\documentclass[superscriptaddress,aps,preprintnumbers,amsmath,showpacs,amssymb,prd,nofootinbib,reprint,onecolumn,floatfix]{revtex4-2}

\usepackage{graphicx}
\usepackage[colorlinks,allcolors=blue]{hyperref} 
\usepackage{amsmath,amssymb,bm}
\usepackage[T1]{fontenc}
\usepackage{mathtools}

\begin{document}

\preprint{TU-1254}

\title{Primordial Black Hole Formation via Inverted Bubble Collapse}

\author{Kai Murai}
\email{kai.murai.e2@tohoku.ac.jp}
\affiliation{Department of Physics, Tohoku University, Sendai, Miyagi 980-8578, Japan}
\author{Kodai Sakurai}
\email{kodai.sakurai.e3@tohoku.ac.jp}
\affiliation{Department of Physics, Tohoku University, Sendai, Miyagi 980-8578, Japan}
\author{Fuminobu Takahashi}
\email{fumi@tohoku.ac.jp}
\affiliation{Department of Physics, Tohoku University, Sendai, Miyagi 980-8578, Japan}

\begin{abstract}
We propose a novel mechanism of primordial black hole (PBH) formation through inverted bubble collapse. In this scenario, bubbles nucleate sparsely in an incomplete first-order phase transition, such that they remain isolated and do not percolate or collide with each other due to the extremely low nucleation rate. This is followed by a bulk phase transition in the rest of the universe that inverts these pre-existing bubbles into false vacuum regions. These spherically symmetric false-vacuum bubbles subsequently collapse to form PBHs. Unlike conventional PBH formation mechanisms associated with domain wall collapse or bubble coalescence, our inverted bubble collapse mechanism naturally ensures spherical collapse. We demonstrate that, when applied to the singlet extension of the Standard Model, this mechanism can produce highly monochromatic PBHs with masses up to ${\cal O}(10^{-7}\,\text{--}\,10^{-5}) M_\odot$, which potentially explain the microlensing events observed in the OGLE and Subaru HSC data.
\end{abstract}

\maketitle

\section{Introduction}

The composition of our universe presents one of the most profound mysteries in modern cosmology. Observational evidence indicates that approximately one-quarter of the energy density of the universe consists of dark matter, whose nature remains unknown~\cite{Planck:2018vyg, Rubin:1970zza, Zwicky:1933gu}. Among various candidates proposed to explain dark matter, primordial black holes (PBHs) stand out as an intriguing possibility~\cite{Hawking:1971vc, Carr:1974nx, Carr:1975qj,Frampton:2010sw, Carr:2016drx}.  While PBHs could potentially account for all of the dark matter, they might also coexist with other dark matter components. Indeed, scenarios in which PBHs constitute a non-negligible fraction or even all of the dark matter have been widely studied.~\cite{Carr:2020gox,Carr:2020xqk, Green:2020jor, Frampton:2010sw}.

The formation of PBHs requires significant density perturbations in the early universe. Many proposed mechanisms rely on superhorizon-scale fluctuations, with one widely studied scenario involving the generation of large curvature perturbations during inflation that later reenter the horizon and collapse into PBHs~\cite{Zeldovich:1967lct,Hawking:1971vc,Carr:1974nx,Carr:1975qj,Yokoyama:1995ex,GarciaBellido1996, Kawasaki1998}. However, such scenarios often require fine-tuning of the inflaton potential, particularly in creating ultra-flat regions in the case of single-field inflation models, which imposes non-trivial constraints on inflationary model building. Another possibility is to make use of the large isocurvature perturbations of baryons or dark matter. For instance, PBH formation from collapsing baryon/axion bubbles represents an interesting realization of such a mechanism~\cite{Dolgov:1992pu,Dolgov:2008wu,Kawasaki:2019iis,Kitajima:2020kig,Kawasaki:2021zir,Kasai:2022vhq,Kasai:2023ofh,Kasai:2023qic,Kasai:2024tgu}.

Alternative mechanisms for generating large overdensities include those involving topological defects such as domain walls~\cite{Ferrer:2018uiu,Ge:2019ihf,Liu:2019lul,Gelmini:2023ngs,Kitajima:2023cek,Ge:2023rrq,Gouttenoire:2023ftk,Gouttenoire:2023gbn,Ferreira:2024eru,Dunsky:2024zdo,Gouttenoire:2025ofv} and first-order phase transitions (FOPTs)~\cite{Sato:1981bf,Kodama:1981gu,Kodama:1982sf,Maeda:1981gw,Jedamzik:1999am,Liu:2021svg,Hashino:2021qoq,Hashino:2022tcs,Kawana:2022olo,Lewicki:2023ioy,Gouttenoire:2023naa,Gouttenoire:2023bqy,Gouttenoire:2023pxh,Balaji:2024rvo,Lewicki:2024ghw,Flores:2024lng,Kanemura:2024pae,Lewicki:2024sfw,Hashino:2025fse}. 
However, these scenarios face fundamental challenges, as the overdense regions are expected to deviate significantly from spherical symmetry, and it remains unclear how much deviation is allowed for successful PBH formation (see Refs.~\cite{Yoo:2020lmg,Escriva:2024aeo,Escriva:2024lmm} for the discussion on the PBH formation from Gaussian curvature perturbations).
These uncertainties make it challenging to predict the PBH abundance and understand the collapse process.

In this paper, we propose a novel mechanism for PBH formation through inverted bubble collapse (IBC) in an incomplete FOPT followed by a bulk phase transition. 
In our scenario, the FOPT is incomplete because the bubble nucleation rate is so low that bubbles remain isolated and the transition never completes through percolation.
These bubbles are spherically symmetric when nucleated~\cite{Coleman:1977th}, and maintain this symmetry unless they collide. Crucially, before this FOPT is completed, a second phase transition occurs throughout the universe except within the bubbles, transforming the bubble interiors into false vacuum regions. Consequently, the initially expanding bubbles begin to contract and eventually collapse. When sufficiently energetic bubbles collapse, they can form PBHs while naturally preserving spherical symmetry. Unlike PBH formation scenarios involving topological defect collapse or bubble coalescence, our mechanism allows for precise prediction of PBH abundance in terms of model parameters thanks to this preserved spherical symmetry. 

We demonstrate that our mechanism can be implemented in the context of the electroweak phase transition (EWPT) by extending the Standard Model (SM) with singlet scalar fields.
In this setup, an initial incomplete FOPT and the subsequent bulk phase transition occur in the singlet sector, followed by the EWPT in the Higgs direction. We show that this scenario can produce highly monochromatic PBHs with masses up to ${\cal O}(10^{-7}\,\text{--}\,10^{-5}) M_\odot$, which could potentially explain the microlensing events observed in the OGLE and Subaru HSC data~\cite{Mroz:2017mvf,Niikura:2017zjd,Niikura:2019kqi} and will also be probed by the ongoing Subaru HSC searches for PBHs and other compact objects.
Depending on the model parameters, it can also generate PBHs with much smaller masses that could explain all dark matter. Furthermore, if either the bulk phase transition in the singlet sector or the EWPT is of first-order, it could generate gravitational waves detectable by future space-borne interferometers such as LISA~\cite{LISA:2022kgy} (see also Refs.~\cite{Hashino:2016xoj,Vaskonen:2016yiu,Beniwal:2017eik,Hashino:2018wee,Ellis:2018mja,Ellis:2018mja,Alves:2018jsw,Alanne:2019bsm,Ellis:2022lft,Ramsey-Musolf:2024ykk} for  theoretical studies of gravitational waves in the singlet extension).

Lastly, we note that the expanding and shrinking bubbles have also been studied in  Ref.~\cite{Ai:2024cka}, which focused on the bubble nucleation during the very early reheating stage.
In their scenario, the shrinking bubbles leave small density perturbations that later collapse to PBHs, which grow via mass accretion during the inflaton matter dominated universe. In contrast, our mechanism operates in the radiation dominated (RD) universe and forms PBHs directly through bubble collapse. Furthermore, our approach allows us to calculate the PBH mass spectrum solely from the phase transition parameters, independent of specific inflation models and reheating processes.

\section{PBH formation in the IBC mechanism}

In this section, we derive the mass distribution of PBHs formed through our IBC mechanism. 
We first describe the outline of our IBC mechanism.
Then, we calculate the bubble size distribution, determine their collapse conditions, and derive the mass spectrum of PBHs.

\subsection{Outline of the IBC mechanism}

\begin{figure*}[t]
    \centering
    \includegraphics[width=.9\textwidth]{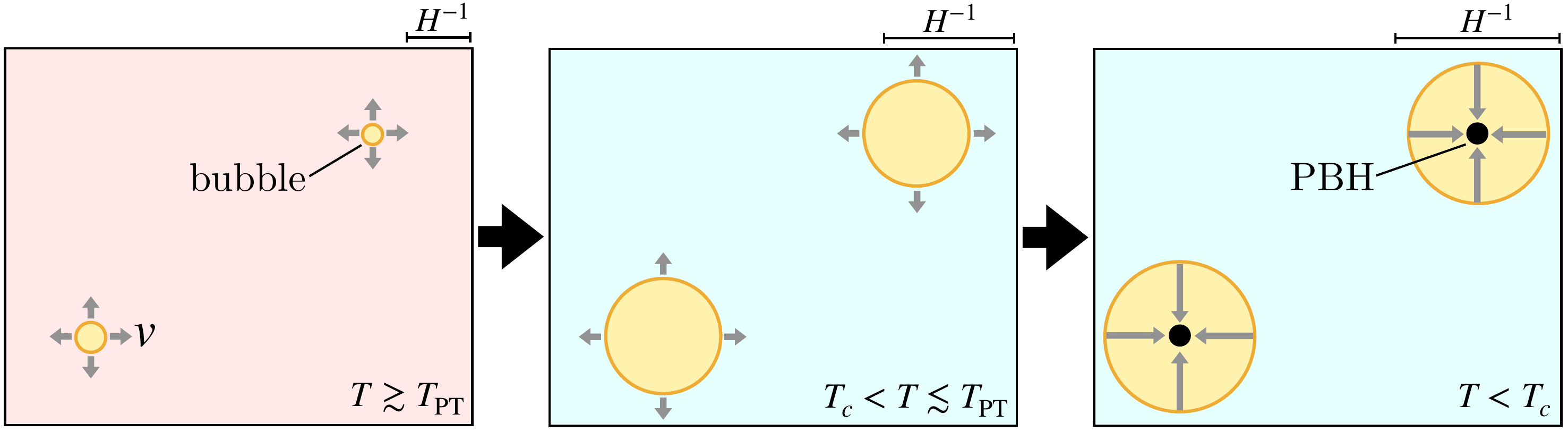}
    \caption{%
    Schematic illustration of the IBC mechanism.
    Left panel: At high temperatures, bubbles nucleate at a low rate, and isolated bubbles expand with velocity $v$.
    Center panel: After the bulk phase transition at $T = T_\mathrm{PT}$, bubble nucleation ceases, while the bubbles continue to expand.
    Right panel: For $T < T_c$, the bubbles shrink and collapse into PBHs.    
    }
    \label{fig: IBC}
\end{figure*}

In our IBC mechanism, an incomplete FOPT is followed by a bulk phase transition at $T = T_\mathrm{PT}$.
At high temperatures, the FOPT proceeds via bubble nucleation.
As long as the nucleation rate per unit volume, $\Gamma(T)$, satisfies $\Gamma(T) \ll H^4$, where $H$ is the Hubble parameter, bubbles of the true vacuum are formed only in a tiny fraction of Hubble volumes, and the FOPT hardly progresses even though each bubble expands with velocity $v$.
Once $\Gamma$ becomes comparable to or larger than $H^4$, bubbles percolate, and the FOPT would normally complete.
In our scenario, we instead consider the situation where 
the bulk phase transition occurs while $\Gamma(T) \ll H^4$, so the FOPT remains incomplete, leaving isolated bubbles with a number density below one bubble per Hubble volume at $T = T_\mathrm{PT}$.
After the bulk phase transition, $\Gamma(T)$ becomes zero or negligibly small, and the bubble nucleation ceases.
As the universe cools down, the bulk region transitions to the true vacuum, while the bubble interiors effectively become regions of false vacuum.
Then, the bubbles start to shrink at $T = T_c$.
If the energy of the false vacuum is efficiently converted into the kinetic energy of the bubble wall, and
 the total energy is concentrated within the Schwarzschild radius, the contracting bubbles collapse into PBHs.
A schematic illustration of the IBC mechanism is shown in Fig.~\ref{fig: IBC}.

\subsection{Bubble size distribution}

To calculate the bubble size distribution, we first need to track the evolution of bubbles in the cooling universe. 
We evaluate this at time $t = t_c$ corresponding to $T = T_c$, when the bubble walls start to shrink after their interiors have become false vacuum regions through the bulk phase transition.
Note that $t_c$ is generically later than the bulk phase transition that occurs at $t=t_\mathrm{PT}$ corresponding to $T = T_\mathrm{PT}$.

Well before the bubble percolation, the number density of the bubbles, $n_{\mathrm{b}}(t)$, satisfies
\begin{align}
    \frac{\mathrm{d} n_\mathrm{b}}{\mathrm{d} t}
    =
    - 3 H n_\mathrm{b} 
    + \Gamma(T(t))
    \ ,
\end{align}
where $\Gamma(T)$ is the bubble nucleation rate per unit volume, which depends on the cosmic temperature $T$ and grows exponentially as the universe cools down.
The solution to this equation is 
\begin{align}
    n_\mathrm{b}(t) a^3 (t)
    =
    e^{\int^t \mathrm{d}t' a^3(t') \Gamma(T(t'))}
    \ .
\end{align}
With this solution for the bubble number density, we can derive the size distribution. A key ingredient is the relation between a bubble's nucleation time $t_n$ and its size $R$ at $t = t_c$. Assuming that each bubble expands with a constant velocity $v$ after nucleation, the size at $t = t_c$ is given by
\begin{align}
    R(t_c; t_n)
    &=
    a(t_c) \left( 
        \frac{R_\mathrm{in}}{a(t_n)} + \int_{t_n}^{t_c} \mathrm{d} \tilde{t} \, \frac{v}{a(\tilde{t})}
    \right)
    \nonumber \\
    &=
    \sqrt{\frac{t_c}{t_n}} R_\mathrm{in}
    + 2 v t_c \left[ 1 - \left(\frac{t_n}{t_c} \right)^{1/2} \right] 
    \ ,
    \label{eq: bubble radius}
\end{align}
where $a$ is the scale factor, and $R_\mathrm{in}$ is the physical critical radius at the bubble nucleation.
In the second equality, we used $a \propto t^{1/2}$, as we focus on PBH formation during the RD era in this paper.
As a result, the size distribution of the bubble at $t = t_c$ is given by 
\begin{align}
    \frac{\mathrm{d} n_\mathrm{b}}{\mathrm{d} R}(t_c)
    &=
    \left| \frac{\mathrm{d} t_R}{\mathrm{d} R(t_c; t_R)} \right|
    \frac{1}{a(t_c)^3}
    \frac{\mathrm{d} n_\mathrm{b} a^3}{\mathrm{d} t}(t_R)
    \nonumber \\
    &=
    \frac{\Gamma(T_R)}{v}
    \frac{t_R^2}{t_c^2}
    \left( 
        1 + 
        \frac{R_\mathrm{in}}{2 v t_R} 
    \right)^{-1}
    \ ,
\end{align}
where $t_R$ is the nucleation time of a bubble whose radius is $R$ at $t = t_c$, i.e., $t_R$ is the time satisfying $R(t_c;t_R) = R$, and we define $T_R \equiv T(t_R)$.
Typically, the contribution of $R_\mathrm{in}$ in Eq.~\eqref{eq: bubble radius} becomes negligibly small soon after bubble nucleation, so it can be dropped.
Then, we obtain
\begin{align}
    t_R 
    \simeq 
    t_c \left( 1 - \frac{R}{2 v t_c} \right)^2
    \ ,
    \label{eq: t vs R}
\end{align}
and
\begin{align}
    \frac{\mathrm{d} n_\mathrm{b}}{\mathrm{d} R}(t_c)
    &\simeq 
    \frac{\Gamma(T(t_R))}{v}
    \left( 1 - \frac{R}{2 v t_c} \right)^{4}
    \ .
\end{align}
Note that the bubble radius is bounded by the sound horizon as
\begin{align}
    R \lesssim 2 v t_c
    \ .
\end{align}

\subsection{PBH formation}

Next, we discuss the PBH formation from a false vacuum bubble.
Here, we denote the potential energy difference between the inside and outside of the bubble by $\Delta V$, which we treat as constant during the relevant timescale of bubble collapse. In realistic situations, $\Delta V$ depends on temperature, as we will see later; however, for simplicity, we treat it as a constant in the following analysis.

Let us assume that the radius of the bubble is $R$ when the bubble wall starts to shrink after its interior transitions to the false vacuum.
Then, the energy of the bubble can be approximately expressed as
\begin{align}
    M_\mathrm{b}
    \simeq 
    \frac{4 \pi}{3} R^3 \Delta V
    \ .
    \label{eq: bubble mass}
\end{align}
Under certain situations, when the energy difference is sufficiently large, the bubble wall can undergo runaway acceleration even in cases with plasma dissipation, whereas in the absence of dissipation, runaway is naturally achieved~\cite{Bodeker:2009qy,Bodeker:2017cim}.
Then, as the bubble shrinks, a part of the vacuum energy swept by the bubble wall is accumulated in the kinetic energy of the wall, and this energy gets progressively concentrated inside a smaller region.
In such runaway cases, we introduce an efficiency parameter $\epsilon (\leq 1)$ to characterize the fraction of vacuum energy converted into the wall's kinetic energy. 
Then, if the bubble size becomes smaller than the Schwarzschild radius, $R_\mathrm{s} = 2  G \epsilon M_\mathrm{b}$, the bubble collapses into a PBH.
Considering that the bubble can be as small as the width of the bubble wall, $\delta$, the condition for the PBH formation is given by%
\footnote{%
In fact, the bubble wall width becomes thinner due to the Lorentz contraction as it gets accelerated. The wall width $\delta$ here is assumed to include this effect.
}  
\begin{align}
    R_\mathrm{s}
    =
    2  G \epsilon M_\mathrm{b}
    \simeq
    \frac{\epsilon R^3 \Delta V}{3 M_\mathrm{Pl}^2} 
    \gtrsim 
    \delta 
    \ .
\end{align}
Considering the upper bound of the bubble radius, we obtain the possible range of the PBH mass as 
\begin{align}
    M_\mathrm{min}
    \equiv 
    \frac{\delta}{2 G}
    \lesssim 
    M
    \lesssim 
    M_\mathrm{max}
    \equiv
    \frac{32 \pi}{3} \epsilon v^3 t_c^3 \Delta V
    \ .
\end{align}
Since the bubble size never exceeds the particle horizon, the false vacuum energy inside the bubble does not dominate the universe before collapsing into a PBH. Thus, $ \Delta V \lesssim 3 M_\mathrm{Pl}^2 H^2 $ holds until the PBH is formed. 
As a result, we obtain the condition for $\Delta V$ given by
\begin{align}
    \frac{3 M_\mathrm{Pl}^2 \delta}{\epsilon R^3}
    \lesssim 
    \Delta V 
    \lesssim 
    3 M_\mathrm{Pl}^2 H_c^2
    =
    \frac{\pi^2 g_*}{30} T_c^4
    \ ,
\end{align}
where $H_c$ denotes the Hubble parameter when $t = t_c$.
In the following, we assume that $M_\mathrm{min}$ is much smaller than the mass region of interest, which is a condition typically satisfied in realistic situations.%
\footnote{%
The width of the bubble wall in the rest frame is typically estimated by the inverse of the mass scale of the field driving the phase transition. For instance, if we take the mass scale to be 100\,GeV, we obtain $M_\mathrm{min} \simeq 6 \times 10^{-22}M_\odot/\gamma$ with $\gamma$ being the Lorentz factor of the wall.}

\subsection{PBH mass function}

Taking into account the size distribution of the bubbles, we obtain the mass function of the PBHs given by
\begin{align}
    \frac{\mathrm{d} \rho_\mathrm{PBH}}{\mathrm{d} \ln M}(t)
    &=
    M
    \frac{\mathrm{d} n_\mathrm{b}}{\mathrm{d} R}(t_c)
    \frac{\mathrm{d} R}{\mathrm{d} \ln M_\mathrm{b}}
    \left( \frac{a(t_c)}{a(t)} \right)^3
    \Theta(M - M_\mathrm{min})
    \nonumber \\
    &=
    \frac{M R}{3} 
    \frac{\Gamma(T_R)}{v}
    \frac{t_R^2}{t_c^2}
    \left( \frac{a(t_c)}{a(t)} \right)^3
    \Theta(M - M_\mathrm{min})
    \ ,
\end{align}
where $R$ is related to $M_\mathrm{b}$ via Eq.~\eqref{eq: bubble mass}, $M = \epsilon M_\mathrm{b}$, and $t$ represents the time after the PBH formation. 

Next, we evaluate the PBH mass function assuming the temperature dependence of the bubble nucleation rate. The bubble nucleation rate is considered to increase progressively toward the bulk phase transition at $T = T_{\rm PT} (\geq T_c)$, and then effectively drops down to zero afterwards.
Thus, around the bulk phase transition, it can be parametrized as%
\footnote{%
We require that the FOPT via bubble nucleation should not occur inside the bubbles, as it would spoil the spherical symmetry and reduce the energy inside them. This condition is met in the case of the singlet extension of the SM studied in the next section.
}
\begin{align}
\label{eq:gamma}
    \Gamma(T)
    \; \simeq \;
    c H_\mathrm{PT}^4\,\, e^{- \alpha \frac{T - T_\mathrm{PT}}{T_\mathrm{PT}}} & \mathrm{~~for~~} T \geq T_\mathrm{PT}
    ,
\end{align}
where $H_\mathrm{PT}$ is the Hubble parameter at the bulk phase transition, and $c$ and $\alpha$ are constant parameters.
From Eqs.~\eqref{eq: t vs R} and \eqref{eq: bubble mass}, we obtain
\begin{align}
    T_R 
    &=
    \sqrt{\frac{t_c}{t_R}} T_c
    \simeq 
    \left( 1 - \frac{R}{2 v t_c} \right)^{-1}
    T_c
    \nonumber \\
    &\simeq 
    \left[ 1 - \frac{H_c}{v} \left( \frac{3 M}{4\pi\epsilon\Delta V} \right)^{1/3} \right]^{-1}
    T_c
    \ .
\end{align}
Then, we obtain the energy ratio to dark matter by
\begin{align}
    \frac{\mathrm{d} f_\mathrm{PBH}}{\mathrm{d} \ln M}
    \equiv &
    \frac{1}{\rho_\mathrm{DM}(t)}
    \frac{\mathrm{d} \rho_\mathrm{PBH}}{\mathrm{d} \ln M}(t)
    \nonumber \\
    \simeq &
    \frac{1}{0.44\,\mathrm{eV}}
    \frac{45}{2 \pi^2 g_{*s}(T_c) T_c^3} 
    \frac{c M H_\mathrm{PT}^4}{3v} 
    \left( \frac{3 M}{4\pi\epsilon\Delta V} \right)^{1/3}
    \left[ 1 - \frac{H_c}{v} \left( \frac{3 M}{4\pi\epsilon\Delta V} \right)^{1/3} \right]^4
    \nonumber \\
    &\times 
    \exp \left[- \alpha \left( 
    \frac{T_c}{T_\mathrm{PT}}
        \left[ 1 - \frac{H_c}{v} \left( \frac{3 M}{4\pi\epsilon\Delta V} \right)^{1/3} \right]^{-1}
        - 1    
    \right) \right]
    \label{eq: fPBH formula}
\end{align}
for $M$ satisfying $M_\mathrm{min} < M < M_\mathrm{max}$ and $T_R > T_\mathrm{PT}$.
Here, we used $\rho_\mathrm{DM}/s = 0.44$\,eV with $s$ being the entropy density.

\begin{figure}[t]
    \centering
    \includegraphics[width=.8\textwidth]{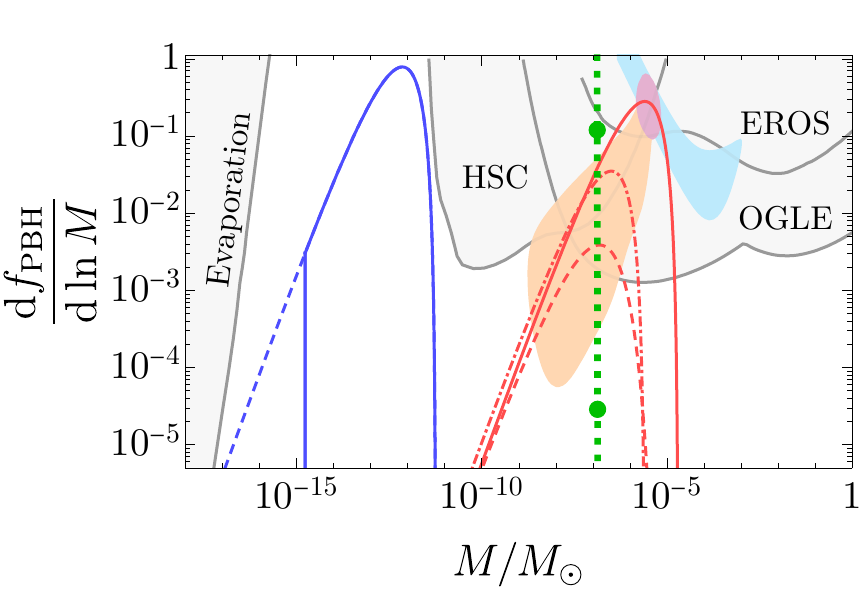}
    \caption{%
        PBH fraction in dark matter, $\mathrm{d}f_\mathrm{PBH}/ \mathrm{d}\ln M$.
        The red and blue solid lines correspond to the parameter sets (a) and (b) in Table.~\ref{tab: params}, respectively. 
        The red dashed and red dot-dashed lines correspond to modifications of parameter set (a), where we set $\alpha = 10$ and $v = 0.5$, respectively.
        For the blue dashed line, we set $T_\mathrm{PT}$ equal to $T_c$ in the parameter set (b) and tune the parameter $c$ so that it overlaps with the blue solid line for larger masses.
        The green dots represent the PBH fraction $f_{\rm PBH}$ in the singlet-extended model, with the dotted line showing its variation with the triple coupling of $S_1$.
        The orange and light blue regions are obtained assuming the lensing event observed in the HSC M31 observation~\cite{Niikura:2017zjd} and OGLE~\cite{Niikura:2019kqi,Mroz:2017mvf} are due to PBHs, respectively.
        The purple region is the joint posterior for them~\cite{Sugiyama:2021xqg}.
        The gray lines represent the observational upper bounds on the PBH abundance by EROS~\cite{EROS-2:2006ryy}, HSC~\cite{Niikura:2017zjd}, OGLE-III+OGLE-IV~\cite{Mroz:2024mse}, OGLE-IV high cadence survey~\cite{Mroz:2024wia}.
        The evaporation constraint is taken from \texttt{PBH bounds}~\cite{Green:2020jor,PBHbounds}.
    }
    \label{fig: fPBH}
\end{figure}

In Fig.~\ref{fig: fPBH}, we show the predicted PBH abundance in our scenario.
We use the parameter sets (a) and (b) shown in Tab.~\ref{tab: params} for the red and blue solid lines, respectively.
These parameter sets correspond to a constant bubble nucleation rate ($\alpha = 0$), fast bubble expansion ($v = 1$), and efficient accumulation of the false vacuum energy ($\epsilon = 1$).
For comparison, we also show the results for the cases with $\alpha = 10$ (red dashed line) and with $v = 0.5$ (red dot-dashed line) based on the parameter set (a).
Note that $v$, $\epsilon$, and $\Delta V$ appear in Eq.~\eqref{eq: fPBH formula} in the form of $v (\epsilon \Delta V)^{1/3}$, and the changes of their values are degenerate in $\mathrm{d} f_\mathrm{PBH}/\mathrm{d} \ln M$.
In these cases, the mass distribution follows $\mathrm{d} f_\mathrm{PBH}/\mathrm{d} \ln M \propto M^{4/3}$ for $M \ll M_\mathrm{max}$ and has a cutoff at $M = M_\mathrm{max}$.
On the other hand, the PBH mass function for the parameter set (b) has a cutoff on smaller mass.
This is because all the bubbles continue to expand at least between $t_\mathrm{PT}$ and $t_c$, and there is a minimal bubble radius at $t = t_c$, leading to the lower bound on the PBH mass.
As we will see below, the PBH mass distribution approaches a monochromatic distribution as $t_c$ is significantly delayed from $t_\mathrm{PT}$.
Setting $T_\mathrm{PT}=T_c$ in parameter set (b) yields the same mass function for a specific choice of $c$, but without the cutoff for small masses (blue dashed line).
\begin{table*}[htbp]
    \caption{Parameter sets for the red and blue solid lines in Fig.~\ref{fig: fPBH}.
    }
    \label{tab: params}
    \begin{tabular}{ c | c c c c c c c}
        Set & $T_\mathrm{PT}$\,[GeV] & $T_c$\,[GeV] & $\Delta V/\rho_\mathrm{tot}(T_c)$ & $v$ & $\epsilon$ & $c$ & $\alpha$
        \\
        \hline \hline
        (a) & 15 & 15 & 0.1 & 1 & 1 & $2.0 \times 10^{-8}$ & 0
        \\
        \hline
        (b) & $3.0 \times 10^4$ & $2.8 \times 10^4$ & 0.1 & 1 & 1 & $1.7 \times 10^{-11}$ & 0
    \end{tabular}
\end{table*}

\section{Singlet Extended Model}

\subsection{Description of the model}
In this section, we apply the IBC mechanism to the singlet extension of the SM where we introduce two real singlets $S_1$ and $S_2$. The scalar potential is given by
\begin{align}
  V(\Phi,S_1,S_2) =&- m_\Phi^2|\Phi|^2+\lambda |\Phi|^4  
  +\mu_{\Phi 1}^{}|\Phi|^2 S_1 
  + \lambda_{\Phi 1} |\Phi|^2 S^2_1+t_1^{}S_1 -m^2_1S^2_1+ \mu_1S^3_1 \nonumber \\
&+ \lambda_1S^4_1-m_2^2S_2^2+\lambda_2S_2^4+\lambda_{\Phi2}|\Phi|^2S_2^2 
+\lambda_{12}S_1^2S_2^2 +\mu_{12}S_1S_2^2,
  \label{Eq:HSM_pot}
\end{align} 
where $\Phi$ denotes the Higgs doublet, and the potential respects the $Z_2$ symmetry acting on  $S_2$.%
\footnote{The collider phenomenology in the extension of the SM with two real singlet scalars with a $Z_2\times Z_2'$ discrete symmetry are studied in Ref.~\cite{Robens:2019kga}.}
The component fields of $\Phi$ and $S_i$ are given by
\begin{align}
\Phi&=
\left[\begin{array}{c}
G^+\\
\frac{1}{\sqrt{2}}\big(v+\phi+iG^0\big)
\end{array}\right]\;, \quad
S_i=v_{i}+s_i\ (i=1,2) \;,
\end{align}
where $v \simeq 246$\,GeV and $v_i$ are the VEVs of $\Phi$ and $S_i$, respectively, and $G^+$ and $G^0$ are Nambu-Goldstone (NG) bosons that are absorbed by the weak gauge bosons.
The shift of $S_1$, $S_1\to S_1+v'_1$, does not change physics~\cite{Kanemura:2016lkz}.
Using this shift transformation, we take $v_1 = 0$ in the
following discussion.
In the mass basis, three CP-even Higgs bosons are identified by the SM-like Higgs boson $(h)$ and heavy Higgs bosons ($H_1$ and $H_2$). 
By setting the tadpole conditions of the scalar potential \eqref{Eq:HSM_pot} and diagonalizing the mass matrix, the original potential parameters can be expressed in terms of physical parameters, i.e., the masses of the Higgs bosons ($m_h$, $m_{H_1}$ and $m_{H_2}$) and the mixing angles $\theta_i$ $(i=1,2,3)$. 
The diagonalization of the mass matrix is presented in Appendix~\ref {app: appendix}. 
We choose the following twelve parameters as input parameters of the model:
\begin{align}
v\;,\quad m_h\;,\quad v_2\;,\quad m_{H_1}\;,\quad m_{H_2}\;,\quad \theta_i\;\ \ (i=1,2,3),\quad  
\mu_{1}\;,\quad \lambda_1\;,\quad \lambda_{\Phi1}\;,\quad \lambda_{12}\;,\quad 
\end{align}
where the EW VEV and the SM-like Higgs boson mass are fixed to be $v=246.2\,{\rm GeV}$ and $m_h=125.09\,{\rm GeV}$.

\subsection{Construction of effective potential}

From the potential~\eqref{Eq:HSM_pot}, one can calculate the effective potential, which consists of the three contributions,
\begin{align}
V_{\rm eff}(\phi,s_i,T)=V_0(\phi,s_i)+V_{\rm CW}(\phi,s_i)+V_T(\phi,s_i,T)~(i=1,2)\;, 
\end{align}
where $V_0(\phi,s_i)$ is the tree-level potential, $V_{\rm CW}(\phi,s_i)$ is Coleman-Weinberg potential up to the one-loop level~\cite{Coleman:1973jx}, 
and $V_T(\phi,s_i,T)$ is the finite-temperature effect of the effective potential~\cite{Dolan:1973qd}. 
The tree-level contribution $V_0$ can be straightforwardly obtained from the potential \eqref{Eq:HSM_pot} with the replacement of $(\Phi,S_i)\to (\phi,s_i)$. 
The remaining parts $V_{\rm CW}$ and $V_T$ depend on the field-dependent masses $\bar{m}^2_k(\phi,s_i)$ of the model.
In terms of the field-dependent masses, $V_{\rm CW}$ and $V_T$ are given by
\begin{align}
V_{\text{CW}} &= \sum_k (-1)^{F_k}  \frac{n_k}{64 \pi^2}  \bar{m}_{k}^4(\phi,s_i) 
\left(  \log{\frac{\bar{m}_{k}^2(\phi,s_i)}{\mu^2_R}}  - c_k   \right) \, , \\
    V_T &= \frac{T^4}{2\pi^2}  \sum_k n_k J_{\mp} \left( \frac{\bar{m}_k(\phi,s_i)}{T} \right) \, ,
\end{align}
where $n_k$ is the number of degrees of freedom of the particle species labeled by $k$, and the constant $c_k$ is 3/2 (5/6) for scalar bosons and fermions (gauge bosons).
Field-dependent masses $\bar{m}_k(\phi,s_i)$ are presented in Appendix~\ref{app:Field-dependent masses}. 
The index $F_k$ is $0\ (1)$ for bosons (fermions). 
The function $J_{\mp}(x)$ is given by
\begin{align}
J_{\mp}(x)  = \pm \int^{\infty}_{0} dy \, y^2 \log\left( 1 \mp e^{-\sqrt{y^2 + x^2}} \right) \, ,
\end{align}
with the minus (plus) sign being for the bosons (fermions). 
Renormalization of the Coleman-Weinberg potential is performed in the $\overline{\rm MS}$ scheme, by which the renormalization scale $\mu_R$ is introduced in the effective potential.\footnote{The Coleman-Weinberg potential can also be evaluated by applying another renormalization scheme, such as on shell-like schemes. The renormalization scheme difference of the effective potential is discussed in Refs.~\cite{Chiang:2018gsn, Athron:2022jyi}.}
In our analysis, we set $\mu_R=m_t$.

We implement the thermal effects of the field-dependent masses by the replacement $\bar{m}^2_k(\phi,s_i)\to \bar{M}^2_k(\phi,s_i,T)$ in the effective potential $V_{\rm eff}$~\cite{Parwani:1991gq}.
The thermally corrected masses $\bar{M}^2_k(\phi,s_i,T)$ involve the temperature dependent self-energy $\Pi_{k}(T)$,
and those for the Higgs field and singlet scalar fields are given by
\begin{align}
    \Pi_{\Phi}(T)&= T^2 \Bigg[\frac{1}{16} \left(3 g^2+g^{\prime 2}\right)+\frac{y_t^2}{4} 
    +\frac{1}{6} \left(\frac{3 \lambda }{2}+\frac{\lambda_{\Phi 1}}{2}+\frac{\lambda_{\Phi 2}}{2}\right)+\frac{\lambda }{4}\Bigg]\;,\\
    \Pi_{S_1}(T)&= T^2 \left[\lambda_{1}+\frac{1}{6} (\lambda_{12}+2 \lambda_{\Phi 1})\right]\;,\\
    \Pi_{S_2}(T)&=T^2 \left[\frac{1}{6} (\lambda_{12}+2 \lambda_{\Phi 2})+\lambda_2\right]\;.
\end{align}
One then obtains a field-dependent mass matrix with the thermal corrections in the basis of $(\phi,s_1,s_2)$: $\bar{{M}}^2=\bar{{\cal M}}^2+{\rm diag}(\Pi_\Phi,\Pi_{S_1},\Pi_{S_2})$, where $\bar{{\cal M}}^2$ is the field-dependent mass matrix without thermal corrections in the basis of $(\phi,s_1,s_2)$ given in Appendix~\ref{app:Field-dependent masses}. 
Diagonalizing $\bar{M}^2$ yields the thermally corrected masses of the CP-even Higgs bosons. 
The thermally corrected masses of the NG boson are obtained by $\bar{M}^2_{G^0,G^\pm}=\bar{m}_{G^0,G^\pm}^2+\Pi_\Phi$. 
The masses of the longitudinal mode of the weak gauge bosons are also replaced with thermally corrected ones~\cite{Carrington:1991hz}.

\subsection{PBH formation}
\begin{figure*}[t]
    \centering
    \includegraphics[width=0.8\linewidth]{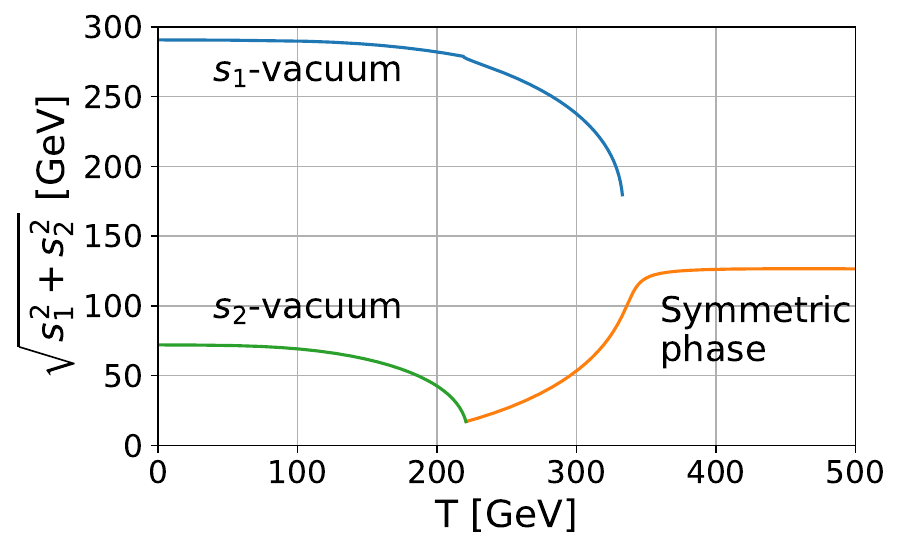}
    \caption{Temperature dependence of three different minima. The blue (green) line shows the local minima appearing in the $S_1$ ($S_2$)-direction ($S_1$ ($S_2$)-vacuum). The orange line shows the minimum that exists at high temperatures.  }
    \label{fig:Phases}
\end{figure*}
\begin{figure*}[t]
    \centering
    \includegraphics[width=1.\linewidth]{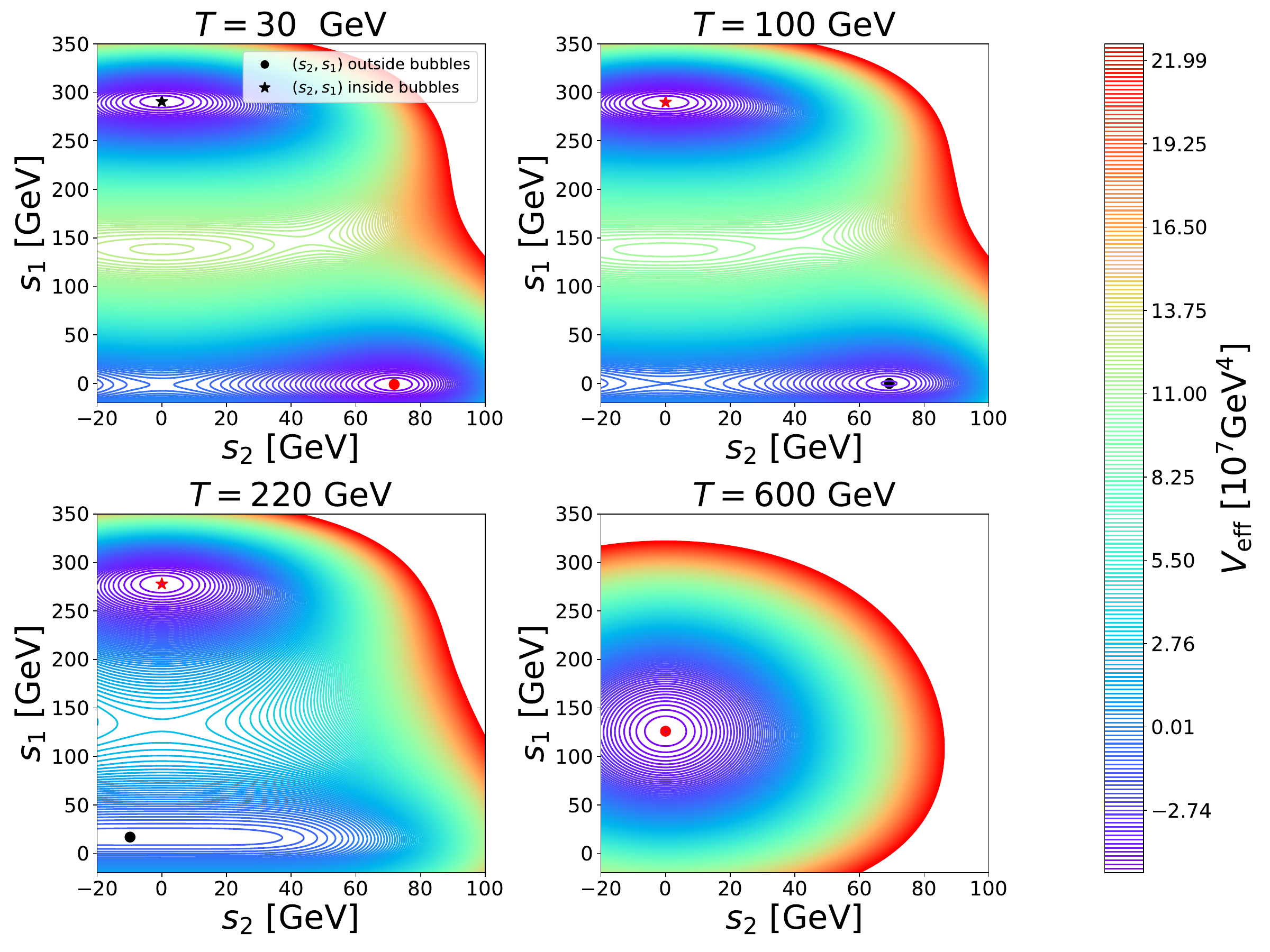}
    \caption{Contours of the effective potential in the $(s_2, s_1)$ plane at temperatures $T=30,\,100,\,220,$ and $600$ GeV. 
    Circle markers indicate the symmetric vacuum or the $s_2$-vacuum, which corresponds to the bulk region, while star markers indicate the $s_1$-vacuum, corresponding to the bubble interior.
    The global minimum at each temperature is shown in red.
    }
    \label{fig:Veff}
\end{figure*}
\begin{figure}[t!]
    \centering \includegraphics[width=.8\textwidth]{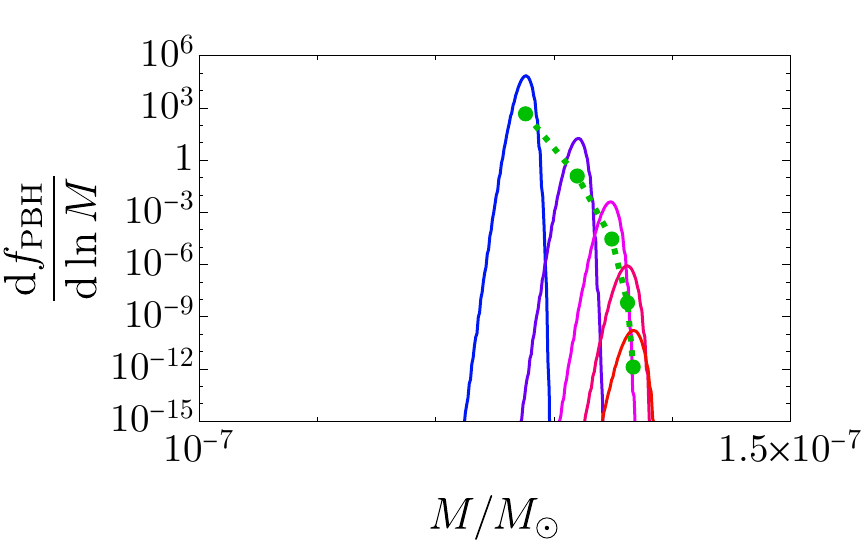}
    \caption{%
       The highly monochromatic PBH mass spectrum in the singlet-extension model. The spectra from top to bottom are calculated with different values of the triple coupling of $S_1$, $\mu_1 = -190.5, -190.49, -190.48, -190.47, -190.46$\,GeV, while other parameters are fixed as \eqref{eq:BP}. Note that the PBH fraction shown by the green dots is about two orders of magnitude smaller than the peak height due to the extremely narrow width.
    }
    \label{fig: fPBH model}
\end{figure}

In our IBC mechanism, the PBH formation is related to phase transitions in the directions of $S_1$ and $S_2$ that occur before the EWPT. 
Therefore, we analyze the scalar potential by setting the CP-even component of $\Phi$  to be $\phi=0$. 
While the required delayed EWSB can be realized for a certain choice of the model parameters~\cite{Ellis:2018mja,Ellis:2022lft},  we focus on the phase transition dynamics in the singlet sector for our purpose.

To realize the IBC mechanism, we set the  input model parameters as follows,
\begin{align}
\label{eq:BP}
&v=246.2\,{\rm GeV},\ m_h=125.09\,{\rm GeV},\ m_{H_1}=223.1\,{\rm GeV},\, m_{H_2}=246\,{\rm GeV},\,v_2=70{\rm GeV},\, \notag \\
&\theta_1=\theta_2=\theta_3=0,\ \lambda_1=0.335,\ \lambda_{12}=0.5,\ \lambda_{\Phi 1}=-0.14,\ \mu_{1}=[-190.49,-190.48]\,{\rm GeV} \;.
\end{align}
For these parameter choices, we have checked that 
the scalar potential satisfies the releavnt theoretical conditions, i.e., perturbative unitarity~\cite{Robens:2019kga} and the stability condition of the potential~\cite{Robens:2019kga}. 
Experimental constraints such as direct searches of the heavy Higgs boson and Higgs coupling measurements for the model with a single real singlet scalar field are discussed using the Run II data of LHC in the Appendix of Ref.~\cite{Ramsey-Musolf:2024ykk}. 
The above parameter points satisfy all these experimental constraints.

There are three relevant local minima that appear at different temperatures. 
This is illustrated in Fig.~\ref{fig:Phases}. 
At sufficiently high temperatures, there is only a unique vacuum where the $Z_2$ symmetry (as well as the EW symmetry) is restored. Note that the singlet $S_1$ has a nonzero VEV in this vacuum, since we do not impose $Z_2$ symmetry on $S_1$.
Then, as the temperature decreases, another local minimum appears along the $S_1$ direction. Let us call this minimum the $S_1$-vacuum. 
We have confirmed that the bubble nucleation rate progressively increases as the universe cools down, where we have used \texttt{CosmoTransitions}~\cite{Wainwright:2011kj} to evaluate the tunneling rate. 
Thus, the $S_1$ field undergoes a FOPT with bubble nucleation, where the $S_1$ field takes a larger value inside the bubbles than outside.
Subsequently, the second PT occurs in the bulk along the $S_2$-direction at $t = t_\mathrm{PT}$, before the completion of the FOPT in the $S_1$ direction.
Let us call this vacuum the $S_2$-vacuum. For the adopted model parameters, the second PT is of second order, and completes around $T_\mathrm{PT} \approx 221$\,GeV.
The energy difference between the $S_1$-vacuum and the $S_2$-vacuum, $\Delta V$, depends on the cosmic temperature. For the adopted parameters, the $S_2$-vacuum is still meta-stable for some time after the second PT, and it becomes the true vacuum around $T_t \approx 60\,\text{--}\,70$\,GeV.
To evaluate $T_c$, we assume that the bubbles start to shrink one Hubble time after the $S_2$-vacuum becomes the true vacuum and use $T_c = T_t/\sqrt{2} \approx 46$\,GeV.
Note that the evolution of the shrinking bubble likely exhibits runaway behavior because the phase transition occurs solely in the singlet sector~\cite{Bodeker:2017cim}.
To illustrate such phase transitions, Fig.~\ref{fig:Veff} shows the effective potential in the $(s_2,s_1)$ plane at different temperatures $T=30,\,100,\,220,$ and $600$\,GeV. 
We set $\mu_1=-190.48\,{\rm GeV}$ while other input parameters are taken as in Eq.~\eqref{eq:BP}. 
One can see that the $S_1$-vacuum first appears in the $S_1$-direction, followed by the appearance of the $S_2$-vacuum in the $S_2$-direction.

When $t_c$ is significantly delayed beyond $t_\mathrm{PT}$, the resultant PBH mass tends to be larger for two reasons: the continued bubble expansion increases the net false vacuum energy inside, and the collapse exhibits more runaway behavior, leading to a larger efficiency parameter $\epsilon$. 
Additionally, as $t_c$ is delayed, the physical size of bubbles asymptotically approaches the sound horizon size at $t=t_c$. This behavior occurs because no new bubbles form after $t_\mathrm{PT}$, and existing bubbles gradually approach the sound horizon. As a consequence, these similarly-sized bubbles collapse to form PBHs with highly monochromatic mass distribution, resulting in a sharp mass peak. 

The PBH abundance shown in Fig.~\ref{fig: fPBH} (represented by green dots and dotted line) corresponds to the parameter set \eqref{eq:BP}.
Due to the extremely narrow mass spectrum in this model, we represent the PBH fraction $f_\mathrm{PBH}$ by green dots in the figure  (using the same numerical scale as $\mathrm{d} f_{\rm PBH}/\mathrm{d} \ln M$ indicated on the vertical axis), with its variation with the singlet triple coupling shown by the dotted lines as we will explain below.
The existence of such viable parameters in this simple setup suggests that our PBH formation mechanism can be naturally realized in a wide class of models with multi-step phase transitions. 

Fig.~\ref{fig: fPBH model} shows the highly monochromatic PBH mass spectrum predicted in the singlet-extension model. It should be noted that due to the extremely narrow peak width, the PBH fraction in dark matter obtained by integrating the distribution (green dots) is about two orders of magnitude smaller than the peak height of the mass distribution.
This spectrum is obtained by substituting the temperature-dependent bubble nucleation rate and $\Delta V$ calculated using \texttt{CosmoTransition} into Eq.~(\ref{eq: fPBH formula}). 
Here, the spectra from top to bottom correspond to different values of the triple coupling of $S_1$, $\mu_1 = -190.5, -190.49, -190.48, -190.47, -190.46$\,GeV, while keeping the other parameters fixed as in Eq.~\eqref{eq:BP}.

\section{Discussion}

For successful PBH formation in our IBC scenario, the scalar potential parameters must be set appropriately; the FOPT should remain incomplete, as bubble collisions would break spherical symmetry and complicate PBH formation.
If PBHs constitute a significant fraction of dark matter, this parameter constraint may be explained by anthropic selection; a completed FOPT would trap our universe in a false vacuum, triggering old inflation and diluting baryons, while without a FOPT, no PBH dark matter would form. 

One of the key elements of our IBC mechanism is the runaway behavior of the collapsing bubbles. There is currently ongoing debate about establishing a definite criterion for the runaway bubble dynamics, but the bubbles in the singlet sector are likely the ones that exhibit the runaway behavior. Even if the bubble dynamics does not exhibit this runaway behavior, we may have successful PBH formation by introducing a short period of inflation~\cite{Yamamoto:1985rd,Lyth:1995ka,Kitajima:2021bjq} after the bubble nucleation. Then, the overdense regions in the interior of the bubbles are stretched beyond the horizon, and they could collapse into PBHs when the bubbles reenter the horizon, if the density perturbations are sizable. 
Note that the late-time inflation does not necessarily have to end inside bubbles, in which case the PBHs can still form when the size of the bubbles becomes smaller than the Hubble radius of the universe outside the bubbles.
This is conceptually related to the PBH formation using vacuum bubbles~\cite{Garriga:2015fdk,Deng:2017uwc} (see also Ref.~\cite{Caravano:2024tlp}).

Furthermore, while this work focuses on pre-EWPT phase transitions in the singlet sector, similar IBC mechanisms might operate during or after the EWPT in singlet or hidden sectors, potentially triggered by the EWPT itself. These scenarios offer additional pathways for PBH formation worthy of future investigation. Also, it is an interesting question if the baryogenesis can be incorporated in our example of the singlet-extended model, especially in a case where the bulk phase transition or the EWPT is of first order.
Investigation of baryogenesis scenarios~\cite{Kuzmin:1985mm,Cohen:1993nk} within this framework is left for future investigation.

\section*{Acknowledgments}
We thank Katsuya Hashino for useful discussions, and Wen-Yuan Ai for informing us of Ref.~\cite{Ai:2024cka}.
This work is supported by JSPS Core-to-Core Program (grant number: JPJSCCA20200002) (F.T.), JSPS KAKENHI Grant Numbers 20H01894 (F.T.), 20H05851 (F.T.), 23KJ0088 (K.M.), 24K17039 (K.M.), and 23KJ0086 (K.S.). 
This article is based upon work from COST Action COSMIC WISPers CA21106, supported by COST (European Cooperation in Science and Technology).

\appendix 

\section{Diagonalization of the mass matrix for the CP-even Higgs boson}
\label{app: appendix}

In this Appendix, we perform the diagonalization of the mass matrix for the CP-even Higgs boson to obtain the relation between the original scalar potential parameter and physical parameters.
The CP-even scalar states $\phi$ and $s_i$ mix with each other. 
The mass matrix in the weak basis $(\phi,s_1,s_2)$ is given by
\begin{align}
{M}^2 &= 
   \left( \begin{array}{ccc}
          {M}_{\phi\phi}^{2} & {M}_{\phi 1}^{2}& {M}_{\phi 2}^{2} \\
          {M}_{\phi 1}^{2} & {M}_{11}^{2}& {M}_{12}^{2} \\
          {M}_{\phi 2}^{2} & {M}_{12}^{2}& {M}_{22}^{2} \\
          \end{array} \right), \\
   M_{\phi\phi}^2 &=  2\lambda v^2,\ \ \ 
   M_{\phi1}^2 =  \mu_{\Phi 1}^{}v,\  \ \ 
   M_{\phi2}^2 =  2 \lambda_{\Phi 2}v v_2, 
   \\
   M_{11}^2 &=  -2m_S^2 + \lambda_{\Phi 1}^{}v^2+2\lambda_{12}v_2^2, \ \ \ 
   M_{12}^2=2\mu_{ 12} v_2,\ \ \ 
   M_{22}^2=8\lambda_{2} v_2^2,
\end{align} 
This mass matrix is diagonalized by the following orthogonal transformation:
\begin{align}
&\begin{pmatrix}  \phi\\ s_1 \\ s_2 \end{pmatrix}=R_{\theta}
\begin{pmatrix}  h\\H_1 \\H_2 \end{pmatrix}, \\
&R_{\theta}=
 \begin{pmatrix}
        c_1 c_2             & -s_1 c_2             & -s_2     \\
        s_1 c_3-c_1 s_2 s_3 & c_1 c_3+ s_1 s_2 s_3 & -c_2 s_3 \\
        c_1 s_2 c_3+s_1 s_3 & c_1 s_3-s_1 s_2 c_3  & c_2 c_3
    \end{pmatrix},
\label{eq:aaa}    
\end{align}
with the shorthanded notations $s_i=\sin\theta_i$, and $c_i=\cos\theta_i$ $(i=1,2,3)$. 
Thus, the masses of $H_i$ and $h$ are obtained by 
\begin{align} \label{eq:mass_diag}
    {\rm diag}(m_{h}^2,m_{H_1}^2,m_{H_2}^2)=R_\theta M^2 R_\theta^T.
\end{align}

Through the relations \eqref{eq:mass_diag} and the tadpole conditions of the scalar potential \eqref{Eq:HSM_pot}, the potential parameters $m_\Phi^2$, $m_i^2\ (i=1,2)$, $t_1$, $\lambda$, $\lambda_{2}$, $\lambda_{\Phi2}$, $\mu_{\Phi 1}$ and $\mu_{12}$ can be expressed in terms of the physical masses and mixing parameters as
\begin{align}
    m_{\Phi}^2=&\lambda  v^2+\lambda_{\Phi 2} v_2^2, 
    \\
    m_1^2=&-\frac{1}{2} m_{h}^2 s^2_{1} c^2_{2}-\frac{1}{2} m_{H_1}^2 (s_{1} s_{2} s_{3}+c_{1} c_{3})^2 
    -\frac{1}{2} m_{H_2}^2 (s_{1} s_{2} c_{3}-c_{1} s_{3})^2+\frac{\lambda_{\Phi 1} v^2}{2}+\lambda_{12} v_2^2, 
    \\
    m_{2}^2=&\frac{\lambda_{\Phi 2} v^2}{2}+2 \lambda_2 v_2^2,
    \\
    t_1=&\frac{1}{2} \left(-\mu_{\Phi 1} v^2-2 \mu_{12} v_2^2\right),
    \\
    \lambda =& \frac{m_{h}^2 c^2_{1} c^2_{2}}{2 v^2}+\frac{m_{H_1}^2 (s_{1} c_{3}-c_{1} s_{2} s_{3})^2}{2 v^2} 
    +\frac{m_{H_2}^2 (c_{1} s_{2} c_{3}+s_{1} s_{3})^2}{2 v^2}, 
    \label{eq:lambda}
    \\
    \lambda_{2}=&\frac{m_{h}^2 s^2_{2}}{8 v_2^2}+\frac{m_{H_1}^2 c^2_{2} s^2_{3}}{8 v_2^2}+\frac{m_{H_2}^2 c^2_{2} c^2_{3}}{8 v_2^2},
    \\
    \lambda_{\Phi 2}=&-\frac{m_{h}^2 c_{1} s_{2} c_{2}}{2 v v_2}+\frac{m_{H_1}^2 c_{2} \left( c_{1} s_{2} s^2_{3}- s_{1} s_{3} c_{3}\right)}{2 v v_2}
    +\frac{m_{H_2}^2 c_{2} \left( c_{1} s_{2} c^2_{3}+ s_{1} s_{3} c_{3}\right)}{2 v v_2},
    \\
    \mu_{\Phi 1}=&-\frac{m_{h}^2 s_{1} c_{1} c^2_{2}}{v} 
    +\frac{m_{H_1}^2}{8 v}\Big[ 2 s_{1} c_{1} \left\{2 c^2_{2}-\left(-s^2_{2}+c^2_{2}-3\right) \left(c^2_{3}-s^2_{3}\right)\right\}-8 s_{2} s_{3} c_{3} \left(c^2_{1}-s^2_{1}\right)\Big]
    \notag \\
    &+\frac{m_{H_2}^2}{8 v} \Big[8 s_{2} s_{3} c_{3} \left(c^2_{1}-s^2_{1}\right) 
    +2 s_{1} c_{1} \left\{\left(-s^2_{2}+c^2_{2}-3\right) \left(c^2_{3}-s^2_{3}\right)+2 c^2_{2}\right\}\Big],
    \\
    \label{eq:mus}
    \mu_{12}=&\frac{m_{h}^2 s_{1} s_{2} c_{2}}{2 v_2}-\frac{m_{H_1}^2 c_{2} \left( s_{1} s_{2} s^2_{3}+ c_{1} s_{3} c_{3}\right)}{2 v_2}
    +\frac{m_{H_2}^2 c_{2} \left( c_{1} s_{3} c_{3}- s_{1} s_{2} c^2_{3}\right)}{2 v_2}\;. 
\end{align}

\section{Field-dependent masses}
\label{app:Field-dependent masses}
In the following, we list field-dependent masses of the scalar bosons in the model with two real singlet scalars. For the CP-even Higgs bosons, from the tree-level effective potential $V_0(\phi,s_i)$, one obtains the field-dependent mass matrix
\begin{align}
\bar{{\cal M}}^2 &= 
   \left( \begin{array}{ccc}
          \bar{{\cal M}}_{\phi\phi}^{2} & \bar{{\cal M}}_{\phi1}^{2}& \bar{{\cal M}}_{\phi2}^{2} \\
          \bar{{\cal M}}_{\phi1}^{2} & \bar{{\cal M}}_{11}^{2}& \bar{{\cal M}}_{12}^{2} \\
          \bar{{\cal M}}_{\phi2}^{2} & \bar{{\cal M}}_{12}^{2}& \bar{{\cal M}}_{22}^{2} \\
          \end{array} \right), \\
    \bar{{\cal M}}^2_{\phi\phi}&=3 \lambda  \phi ^2-m_\Phi^2+\lambda_{\Phi 1} s_1^2+\mu_{\Phi 1} s_1+\lambda_{\Phi 2} s_2^2,\\
\bar{{\cal M}}^2_{\phi1}&=  \phi  (\mu_{\Phi 1}+2 \lambda_{\Phi 1} s_1),\ \ \ 
\bar{{\cal M}}^2_{\phi2}= 2 \lambda_{\Phi 2} s_2 \phi,\\
\bar{{\cal M}}_{11}^2&=\lambda_{\Phi 1} \phi ^2-2 m_S^2+12 \lambda_{1} s_1^2 
+6 \mu_S s_1+2 \lambda_{12} s_2^2, \\
\bar{{\cal M}}_{12}^2&=2 s_2 (\mu_{12}+2 \lambda_{12} s_1), \\
\bar{{\cal M}}_{22}^2&=\lambda_{\Phi 2} \phi ^2-2 m_2^2+2 \lambda_{12} s_1^2 
+2 \mu_{12} s_1+12 \lambda_2 s_2^2. 
\end{align}
Diagonalizing it, one obtains field-dependent masses of the CP-even Higgs bosons, i.e., $\bar{{m}}_h^2,\bar{{m}}_{H_1}^2,\bar{{ m}}_{H_2}^2$. 
The field-dependent masses of the NG bosons are given by
\begin{align}
\bar{m}_{G^0}^2&=
    \lambda \phi^2 -m_\Phi^2+\lambda_{\Phi 1} s_1^2+\mu_{\Phi 1} s_1+\lambda_{\Phi 2} s_2^2, \\
    \bar{m}_{G^\pm}^2&=\bar{m}_{G^0}^2. 
\end{align}
Those of weak gauge bosons and fermions are the same as the SM~\cite{Carrington:1991hz}. 

\bibliographystyle{apsrev4-1}
\bibliography{ref}

\end{document}